\documentclass[submission,copyright,creativecommons]{eptcs}
\pdfoutput=1
\providecommand{\event}{ACL2 Workshop 2015} 

\usepackage{paralist,array,amsmath,amsfonts,amssymb,stmaryrd,hyperref,relsize,extarrows,centernot,esvect}
\usepackage{color,booktabs,fixltx2e,graphicx,subfig,etex,etoolbox,textcomp}
\usepackage{alltt}

\newtheorem{definition}{Definition}

\newcommand\mmit[1]{\ensuremath{\mathit{#1}}}

\def\IEEEQEDclosed{\mbox{\rule[0pt]{1.3ex}{1.3ex}}} 
\def\endIEEEproof{\hspace*{\fill}~\IEEEQEDclosed\par\vspace{5pt}}

\newcommand{\La}[0]{\ensuremath{\langle}}
\newcommand{\Ra}[0]{\ensuremath{\rangle}}

\newcommand{\pfEnd} {\endIEEEproof}
\newcommand{\rankt} {\ensuremath{\mathit{rankt}}}
\newcommand{\rankl} {\ensuremath{\mathit{rankl}}}

\newcommand{\p}[1]{\begin{normalfont}\frenchspacing\texttt{#1}\end{normalfont}}

\newcommand{\comment}[1]{}
\newcommand{\ie}[0]{\emph{i.e.}, }
\newcommand{\eg}[0]{\emph{e.g.}, }

\newcommand\Comment[1]{}

\newcommand{\trans}[0]{\ensuremath{ \rightarrow }}

  \mathchardef \mhyphen="2D

\newcommand{\trs}[1]{\mbox{\ensuremath{\mathcal{M_{\text{#1}}}= \langle S_{#1}, \xrightarrow{#1}, L_{#1} \rangle}}}
\newcommand{\labf}{\ensuremath{\mathcal{L}}}
\newcommand{\M}[1]{\ensuremath{\mathcal{M}_{#1}}}
\newcommand{\Mp}[1]{\ensuremath{\mathcal{M}'}}
\makeatletter
\newcommand{\xrightarrowc}[2][]{\ext@arrow 0359\rightarrowfill@{#1}{\hspace{-4pt}#2}}
\makeatother

\makeatletter
\newdimen\arrow@ht
\setbox\@tempboxa\hbox{\(\xrightarrow{}\)}
\arrow@ht\ht\@tempboxa
\newdimen\plus@wd
\setbox\@tempboxa\hbox{\(\scriptstyle +\)}
\plus@wd\wd\@tempboxa
\def\righttransarrowfill@{\arrowfill@\relbar\relbar{\raisebox{0pt}[\arrow@ht][0pt]{\(\xrightarrow{}^+\hskip-\plus@wd\)}}}
\setbox\@tempboxa\hbox{\(\xrightarrow{}\)}
\arrow@ht\ht\@tempboxa
\newcommand{\tto}[2][]{\ext@arrow 0359\righttransarrowfill@{#1}{#2}\hskip\plus@wd}
\makeatother

\newcommand{\disjtrs}[3]{\ensuremath{\langle S_{#1} \uplus S_{#2},
\xrightarrow{#1} \uplus \xrightarrow{#2}, \labf{#3} \rangle }}

\newcommand{\vbar}{\ensuremath{\ |\ } }

\let\orgautoref\autoref %

\renewcommand{\autoref}[1] {%
  \def\equationautorefname{Eq.}%
  \def\figureautorefname{Fig.}%
  \def\subfigureautorefname{Fig.}%
  \def\subfigureautorefname{Fig.}%
  \def\definitionautorefname{Definition}%
  \orgautoref{#1}%
}

\newcommand{\from}{\ensuremath{\colon}}
\newcommand{\ra}{\ensuremath{\rightarrow\;}}

\newcommand{\xrex}[1]{\ensuremath{\xrightarrow{\;#1\;}}}
\newcommand{\sxrex}[2]{\ensuremath{\xrex{#1}\hspace{-3pt}^{\scalebox{.6}{$<#2$}}}}

\newcommand{\sem}[1]{[ \! [ #1 ] \! ]}

\newcommand{\doIR}[3]{
    \frac{
        \begin{array}{#3} #1 \end{array}
    }{
        \begin{array}{#3} #2 \end{array}
    }
}
\newcommand{\Rule}[2]{\[ \doIR{#1}{#2}{c} \]}

\newcommand{\tuple}[3]{\ensuremath{\langle {#1\ }{#2\ } {#3}\rangle}}
\newcommand{\tuplec}[3]{\ensuremath{\langle {#1},{#2}, {#3}\rangle}}
\newcommand{\qtuplec}[4]{\ensuremath{\langle {#1},{#2},{#3},{#4}\rangle}}

\usepackage{etex,etoolbox}
\makeatletter
\providecommand{\@fourthoffour}[4]{#4}
\def\fixstatement#1{%
  \AtEndEnvironment{#1}{%
    \xdef\pat@label{\expandafter\expandafter\expandafter
      \@fourthoffour\csname#1\endcsname\space\@currentlabel}}}

\globtoksblk\prooftoks{1000}
\newcounter{proofcount}

\long\def\proofatend#1\endproofatend{%
  \edef\next{\noexpand[Proof of \pat@label] }%
  \toks\numexpr\prooftoks+\value{proofcount}\relax=\expandafter{\next#1\pfEnd}
  \stepcounter{proofcount}}

\def\printproofs{%
  \count@=\z@
  \loop
    \the\toks\numexpr\prooftoks+\count@\relax
    \ifnum\count@<\value{proofcount}%
    \advance\count@\@ne
  \repeat}
\makeatother

\DeclareFontFamily{OT1}{bbm}{}
\DeclareFontShape{OT1}{bbm}{m}{n}{%
      <5> <6> <7> <8> <9> <10> <12> <17> gen * bbm%
      <10-12>bbm10%
      <12-17>bbm12%
      <17->bbm17%
      <-5>bbm5}{}

\DeclareSymbolFont{bm}{OT1}{bbm}{m}{n}
\DeclareMathSymbol{\N}{7}{bm}{'116}

\title{Proving Skipping Refinement with ACL2s}%
\author{Mitesh Jain \qquad Panagiotis Manolios%
  \institute{Northeastern University}
  \email{\{jmitesh,pete\}@ccs.neu.edu} \thanks{This research was
    supported in part by DARPA under AFRL Cooperative Agreement
    No.~FA8750-10-2-0233, by NSF grants CCF-1117184 and CCF-1319580,
    and by OSD under contract FA8750-14-C-0024.}}

\def\titlerunning{Proving Skipping Refinement with ACL2s}
\def\authorrunning{Mitesh Jain \& Panagiotis Manolios}

\begin{document}
\maketitle

\begin{abstract}
  We describe three case studies illustrating the use of ACL2s to
  prove the correctness of optimized reactive systems using skipping
  refinement. Reasoning about reactive systems using refinement
  involves defining an abstract, high-level \emph{specification}
  system and a concrete, low-level \emph{implementation} system. Next,
  one shows that the behaviors of the implementation system are
  allowed by the specification system. Skipping refinement allows us
  to reason about implementation systems that can ``skip''
  specification states due to optimizations that allow the
  implementation system to take several specification steps at
  once. Skipping refinement also allows implementation systems to
  \emph{stutter}, \ie to take several steps before completing a
  specification step. We show how ACL2s can be used to prove skipping
  refinement theorems by modeling and proving the correctness of three
  systems: a JVM-inspired stack machine, a simple memory controller,
  and a scalar to vector compiler transformation.
\end{abstract}

\section{Introduction}

Refinement is a powerful method for reasoning about reactive systems.
The idea is that a simple high-level abstract system acts as a
specification for a low-level implementation of a concrete system. The
goal is then to prove that all observable behaviors of the concrete
system are behaviors of the abstract system. It is often the case that
the concrete system requires several steps to match one high-level
step of the abstract system, a phenomenon commonly known as
stuttering. Therefore, notions of refinement usually directly account
for
stuttering~\cite{abadi1991existence,browne1988characterizing,van1990linear}.
However, in the course of engineering an efficient implementation, it
is often the case that a single step of the concrete system can
correspond to several steps of the abstract system, a phenomenon that
is dual of stuttering. For example, in order to reduce memory latency
and effectively utilize memory bandwidth, memory controllers often
buffer requests to memory. The pending requests in the buffer are
analyzed for address locality and then at some time in the future,
multiple locations in the memory are read and updated
simultaneously. Similarly, to improve instruction throughput,
superscalar processors fetch multiple instructions in a single clock
cycle.  These instructions are analyzed for instruction-level
parallelism (\eg the absence of data dependencies), and where possible
multiple instructions are executed in parallel, retired in a single
clock cycle. In both the above examples, updating multiple locations
in memory and retiring multiple instructions in a single clock cycle,
results in scenario where a single step in the optimized
implementation may correspond to multiple steps in the abstract
system. A notion of refinement that only account for stuttering is
therefore not appropriate for reasoning about such optimized systems.

In our companion paper~\cite{manolios2015sks}, we proposed
\emph{skipping refinement}, a new notion of correctness for reasoning
about optimized reactive systems and a proof method that is amenable
for mechanical reasoning. The applicability of skipping refinement was
shown using three case studies: a JVM-inspired stack machine, an
optimized memory controller, and a vectorizing compiler
transformation. In~\cite{manolios2015sks} we focused on finite-state
models for the systems in the first two case studies and used
model-checkers to verify skipping refinement. In this paper, we
consider their corresponding infinite-state models and prove their
correctness in ACL2s, an interactive theorem prover~\cite{acl2s11}. We
also discuss in detail the modeling of vectorizing compiler
transformation and its proof of correctness. In
Section~\ref{sec:motivation}, we motivate the need for a new notion of
refinement with an example. In Section~\ref{sec:sks}, we define
well-founded skipping simulation. In Section~\ref{sec:casestudies} we
discuss the three case studies. We end the paper with conclusion and
future work in Section~\ref{sec:conclusion}.  \vspace{-.2cm}

\section{Motivating Example}
\label{sec:motivation}

To illustrate the notion of skipping simulation, we consider an
example of a discrete-time event simulation (DES)
system~\cite{manolios2015sks}. An abstract high-level specification of
DES is described as follows. Let $E$ be set of events and $V$ be set
of variables. Then a state of abstract DES is a three-tuple
$\langle t, Sch, A\rangle$, where $t$ is a natural number denoting
current time; $Sch$ is a set of pairs $(e, t_e)$, where $e$ is an
event scheduled to be executed at time $t_e \geq t$; $A$ is an
assignment to variables in $V$. The transition relation for the
abstract DES system is defined as follows. If at time $t$ there is no
$(e,t) \in Sch$, \ie there is no event scheduled to be executed at
time $t$, then $t$ is incremented by 1. Else, we
(nondeterministically) choose and execute an event of the form
$(e,t) \in Sch$. The execution of event may result in modifying $A$
and also adding finite number of new pairs $(e',t')$ in $Sch$.  We
require that $t' > t$. Finally execution involves removing the
executed event $(e,t)$ from $Sch$.

Now, consider an optimized, concrete implementation of the abstract
DES system. As before, a state of the concrete system is a three-tuple
$\langle t, Sch, A \rangle$. However, unlike the abstract system which
just increments time by 1 when no events are scheduled for the current
time, the optimized system uses a priority queue to find the next
event to execute. The transition relation is defined as follows. An
event $(e, t_e)$ with the minimum time is selected, $t$ is updated to
$t_e$ and the event $e$ is executed, as in the abstract DES.

Notice that when no events are scheduled for execution at the current
time, the optimized implementation of the discrete-time event
simulation system can run faster than the abstract specification
system by \emph{skipping} over abstract states. This is not a
stuttering step as it results in an observable change in the state of
the concrete DES system ($t$ is update to $t_e$). Also, it does not
correspond to a single step of the specification. Therefore, it is not
possible to prove that the implementation \emph{refines} the
specification using notions of refinement that only allow
stuttering~\cite{abadi1991existence,manolios2003compositional},
because that just is not true. But, intuitively, there is a sense in
which the optimized DES system \emph{does} refine the abstract DES
system. The notion of skipping refinement proposed
in~\cite{manolios2015sks} is an appropriate notion to relate such
systems: a low-level implementation that can run slower (stutter) or
run faster (skip) than the high-level specification.

\vspace{-.2cm}

\section{Skipping Refinement}
\label{sec:sks}

In this section, we first present the notion of well-founded skipping
simulation~\cite{manolios2015sks}. The notion is defined in the
general setting of labeled transition systems (TS) where labeling is
on states. \footnote{Note that labeled transition system are also used
in the literature to refer to transition systems where transitions
(edges) are labeled. However, we prefer to work with TS where states
are labeled.} We also place no restriction on the state space sizes
and the branching factor, and both can be of arbitrary infinite
cardinalities. The generality of TS is useful to model systems that
may exhibit unbounded nondeterminism, for example, modeling a program
in a language with random assignment command $x = ?$, which sets x to
an arbitrary integer~\cite{apt1986countable}.

We first describe the notational conventions used in the
paper. Function application is sometimes denoted by an infix dot
``$.$'' and is left-associative. For a binary relation $R$, we often
use the infix notation $xRy$ instead of $(x, y) \in R$.  The
composition of relation $R$ with itself $i$ times (for $0 < i \leq
\omega$) is denoted $R^i$ ($\omega = \N$ and is the first infinite
ordinal).  Given a relation $R$ and $1<k\leq \omega$, $R^{<k}$ denotes
$\bigcup_{1 \leq i < k} R^i$ and $R^{\geq k}$ denotes $\bigcup_{\omega
> i \geq k} R^i$ .  Instead of $R^{< \omega}$ we often write the more
common $R^+$.  $\uplus$ denotes the disjoint union operator.
Quantified expressions are written as $\langle \emph{Q}x \from r \from
p \rangle$, where \emph{Q} is the quantifier (\eg $\exists, \forall$),
$x$ is the bound variable, $r$ is an expression that denotes the range
of \emph{x} (\emph{true}, if omitted), and $p$ is the body of the
quantifier.

\begin{definition}
  A labeled transition system (TS) is a structure \mbox{$\langle S,
    \ra,L\rangle$}, where $S$ is a non-empty (possibly infinite) set
  of states, $\ra \ \subseteq S \times S$ is a left-total transition
  relation (every state has a successor), and $L$ is a labeling
  function: its domain is $S$ and it tells us what is observable at a
  state.

\end{definition}

Skipping refinement is defined based on well-founded skipping
simulation, a notion that is amenable for mechanical reasoning. This
notion allows us to reason about skipping refinement by checking
mostly local properties, \ie properties involving states and their
successors.  The intuition is, for any pair of states $s, w$, which
are related and a state $u$ such that $s \xrightarrow{} u$, there are
four cases to consider (\autoref{fig:wfsk}): (a) either we can match
the move from $s$ to $u$ right away, \ie there is a $v$ such that
$w \xrightarrow{} v$ and $u$ is related to $v$, or (b) there is
stuttering on the left, or (c) there is stuttering on the right, or
(d) there is skipping on the left.

\vspace{-8cm}
\begin{minipage}[c]{\textwidth}
\hspace{-.8cm} %
\includegraphics[scale=.7]{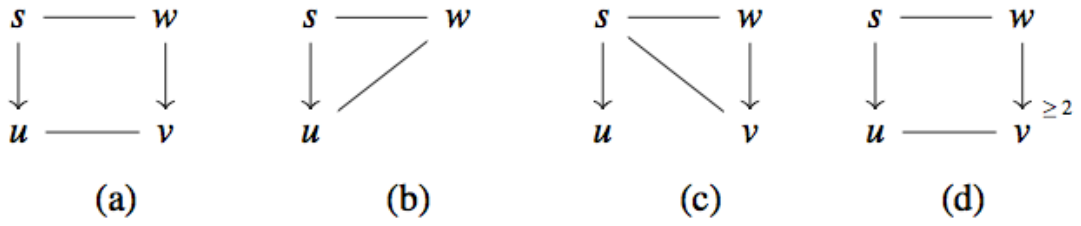}
\label{fig:wfsk}
\end{minipage}
\vspace{-8.5cm}

\begin{definition} [Well-founded Skipping]
  \label{def:wfsk}
  $B \subseteq S \times S $ is a well-founded skipping relation on a
  transition system {\trs{}} iff: {
	\begin{enumerate}
	\item [(WFSK1)] $\langle \forall s,w \in S \from s B w \from L.s =
	  L.w\rangle$
	\item [(WFSK2)] There exist functions, $\rankt\from S\times S
	  \rightarrow W$, $\rankl \from S \times S \times S
	  \rightarrow \omega$, such that $\langle W, \prec \rangle$ is
	  well-founded and 
	  {\setlength{\abovedisplayskip}{3pt}
	    \setlength{\belowdisplayskip}{5pt}
	    \begin{flalign*}
              \La \forall s, & u,w \in S: s \xrightarrow{} u
              \wedge sBw: & \\
	      &\text{(a) } \langle \exists v \from w \xrightarrow{} v\from uBv \rangle \ \vee&\\
	      &\text{(b) } (uBw \wedge \rankt(u,w) \prec
	      \rankt(s,w)) \ \vee&\\
	      &\text{(c) } \langle \exists v \from w \xrightarrow{} v
	      \from s B v \wedge \rankl(v,s,u) < \rankl(w,s,u) \Ra \ \vee&\\
	      &\text{(d) } \langle \exists v : w \rightarrow^{\geq 2} v
	      \from uBv \Ra \Ra
	    \end{flalign*}
	  }
	\end{enumerate}
      }
\end{definition}

In the above definition, conditions (WFSK2a) to (WFSK2c) require
reasoning only about single step $\ra$ of the transition system. But
condition (WFSK2d) requires us to check that there exists a $v$ such
that $v$ is \emph{reachable} from $w$ in two or more steps and $uBv$
holds. Reasoning about reachability, in general, is not
local. However, for the kinds of optimized systems we are interested
in the number of abstract steps that a concrete step corresponds to is
bounded by a constant---a bound determined early on in the design
phase. For example, the maximum number of abstract steps that a
concrete step of a superscalar processor can correspond to is the
number of instruction that the designer decides to retire in a single
cycle. This is a constant that is decided early on in the design
phase.  Therefore, for such systems we can still reason using
``local'' methods. Furthermore, in the case this constant is a
``small'' number, condition (WFKS2d) can be checked by simply
unrolling the transition relation of the concrete system, an
observation that we exploit in our first two case studies. On the
other hand, this simplification is not always possible. For example,
in the optimized DES system describe above, notice that number of
abstract steps that optimized DES can take corresponds to the
difference between current time and earliest time an event is
scheduled for execution. This difference can not be a priori bounded
by a constant.

We now define the notion of skipping refinement, a notion that relates
\emph{two} transition systems: an \emph{abstract} transition system
and a \emph{concrete} transition system. In order to define skipping
refinement, we make use of \emph{refinement maps}, functions that map
states of the concrete system to states of the abstract
system. Informally, if the concrete system is a skipping refinement of
the abstract system then its observable behaviors are also behavior of
the abstract system modulo skipping (which includes stuttering). For
example, in our running example of DES, if the refinement map is the
identity function then it is easy to see that any behavior of the
optimized system is a behavior of the abstract system modulo
skipping. In practice, the abstract system and the concrete system are
described at different levels of abstraction. Refinement maps along
with the labeling function enable us to define what is observable at
concrete states from the view point of the abstract system.

\begin{definition}[Skipping Refinement]
  \label{def:skipref}
  Let \trs{A} and \trs{C} be transition systems and let $\mathit {r
  \from S_C \ra S_A}$ be a \emph{refinement map}.
  We say $\M{C}$ is a \textit{skipping refinement} of $\M{A}$ 
  with respect to $r$, written
  $\M{C} \lesssim_r \M{A}$, if there exists a relation $B \subseteq
  S_C \times S_A$ such that all of the following hold.
  \begin{enumerate}
  \item $\langle \forall s \in S_C :: sBr.s\rangle $ \emph{and}
  \item B is an WFSK on \disjtrs{C}{A}{} where $\labf.s = L_A(s)$ for
    $s \in S_A$, and $\labf.s = L_A(r.s)$ for $s \in S_C$.
  \end{enumerate}
\end{definition}

Well-founded skipping gives us a simple proof rule to determine if a
concrete transition system \M{C} is a skipping refinement of an
abstract transition system \M{A} with respect to a refinement map
$r$. Given a refinement map $r : S_C \rightarrow S_A$ and relation
$B \subseteq S_C \times S_A$, we check the following two conditions:
(a) for all $s \in S_C$, $sBr.s$ and (b) if $B$ is a WFSK on the
disjoint union of \M{C} and \M{A}. If (a) and (b) hold,
$\M{C} \lesssim_r \M{A}$. For a more detailed exposition of skipping
refinement we refer the reader to our companion
paper~\cite{manolios2015sks}.

Notice that we place no restrictions on refinement maps. When
refinement is used in specific contexts it is often useful to place
restrictions on what a refinement map can do, \eg we may require for
every $s \in S_C$ that $L_A(r.s)$ is a projection of $L_C(s)$. The
generality of refinement map is useful in all three case studies
considered in this paper, where a simple refinement map that is a
projection function would not have sufficed. 
\vspace{-.2cm}

\section{Case Studies}
\label{sec:casestudies}

We consider three case studies. The first case study is a hardware
implementation of a JVM-inspired stack machine with an instruction
buffer. The second case study is a memory controller with an
optimization that eliminates redundant writes to memory. The third
case study is a compiler transformation that vectorizes a list of
scalar instructions. For each case study we model the abstract system
and the concrete system in ACL2s. We define an appropriate refinement
map and prove that the implementation refines the specification using
well-founded skipping simulation.

We first briefly list some conventions used to describe the syntax and
the semantics of the systems. Adding element \mmit{e} to the beginning
or end of a list (or an array) \mmit{l} is denoted by $e \, {::} \, l$
and $l \,{::}\, e$, respectively. Each transition consists of a
\mmit{state:}{\mmit{condition_1},\ldots,\mmit{condition_n}} pair above a
line, followed by the next state below the line. If a concrete state
matches the state in a transition and satisfies each of the
conditions, then the state can transition to the state below the
line. We formalize the operational semantics of the machines by
describing the effect of each instruction on the state of the
machine. The proof scripts are publicly available~\cite{sksmodel}.

\subsection{JVM-inspired Stack Machine}
\label{sec:stackmachine}

In this case study, we verify a stack machine inspired by the Java
Virtual Machine (JVM). Java processors have been proposed as an
alternative to just-in-time compilers to improve the performance of
Java programs. Java processors such as JME~\cite{hardin2001real} fetch
bytecodes from an instruction memory and store them in an instruction
buffer. The bytecodes in the b{\rm }uffer are analyzed to perform
instruction-level optimizations \eg instruction folding. In this case
study, we verify BSTK, a simple hardware implementation of part of the
JVM. BSTK is an incomplete and inaccurate model of JVM that only
models an instruction memory, an instruction buffer and a stack. Only
a small subset of JVM instructions are supported (\p{push},
\p{pop}, \p{top}, \p{nop}). However, even such a simple model
is sufficient to exhibit the applicability of skipping simulation and
the limitations of current hardware model-checking tools.

STK is the high-level specification with respect to which we verify
the correctness of BSTK, the implementation. Their behaviors are
defined using abstract transition systems.  The syntax and the
operational semantics are shown in \autoref{fig:stacksemantics}.

The state of STK consists of an instruction memory \p{imem}; a program
counter \p{pc}; and a stack \p{stk}. An instruction is one of \p{push,
pop, top}, and \p{nop}. We use the \p{listof} combinator in
\p{defdata} to encode the instruction memory as list of instructions
and stack as a list of elements~\cite{defdata}. The program counter is
encoded as a natural number using the primitive data type \p{nat}. We
then compound these components to encode state of STK using the
\p{defdata} \p{record} construct. The \p{defdata} framework introduces
a constructor function \p{sstate}, a set of accessor functions for
each field (\eg \p{sstate-imem}), a recognizer function \p{sstatep}
identifying the state, an enumerator \p{nth-sstate} and several useful
theorems to reason about compositions of these functions.

\begin{alltt}
(defdata el all)

(defdata stack (listof el))

(defdata inst (oneof (list 'pop) (list 'top)
                     (list 'nop) (list 'push el)))

(defdata inst-mem (listof inst))

(defdata sstate (record (imem . inst-mem)
                        (pc . nat)
                        (stk . stack)))
\end{alltt}


\begin{figure}[p!]
\begin{minipage}{\textwidth}
 \begin{flalign*}
   &\mathit{stk} := \mathit{[] \vbar el\,{::}\,stk}&\text{(Stack)}&\\
   &\mathit{inst} := \mathit{\langle push\ e \rangle \vbar \langle pop
     \rangle \vbar \langle top \rangle \vbar \langle nop\rangle}&\text{(Instruction)}&\\
   &\mathit{imem} := \mathit{[] \vbar inst\,{::}\,imem}&\text{(Program)}&\\
   &\mathit{pc}:= 0 \vbar 1 \vbar \cdots \vbar n \vbar \cdots &\text{(Program Counter)}&\\
   &\mathit{ibuf}:= \mathit{[inst_1, \ldots , inst_k]} &\text{(Instruction Buffer)}&\\
   &\mathit{sstate} := \mathit{\tuplec{imem}{pc}{stk}}&\text{(STK State)}&\\
   &\mathit{istate} := \mathit{\langle imem, pc, ibuf, stk\rangle}
   &\text{(BSTK State)}&\qquad
 \end{flalign*}
 \end{minipage}

\fbox{\begin{minipage}{.978\textwidth}
 STK ($\xrightarrow{A}$) where $s=$ capacity of \mmit{stk}, $t=\mmit{|stk|}$ %
 \Rule{%
   \mathit{\tuplec{imem}{pc}{stk} : imem[pc] =\langle push\
     v\rangle, t<s}%
 }%
 {%
   \mathit{\tuplec{imem}{pc+1}{v\,{::}\,stk}}%
 }%
 \Rule{%
   \mathit{\tuplec{imem}{pc}{stk} : imem[pc] =\langle push\
     v\rangle, t=s}%
 }%
 {%
   \mathit{\tuplec{imem}{pc+1}{stk}}%
 }%
 \Rule{%
   \mathit{\tuplec{imem}{pc}{[]} : imem[pc] = \langle pop \rangle}%
 }%
 {%
   \mathit{\tuplec{imem}{pc+1}{[]}}%
 }%
 \Rule{%
   \mathit{\tuplec{imem}{pc}{v\,{::}\,stk} : imem[pc] = \langle pop \rangle}%
 }%
 {%
   \mathit{\tuplec{imem}{pc+1}{stk}}%
 }%
 \Rule{%
   \mathit{\tuplec{imem}{pc}{stk} : imem[pc] = \langle top \rangle}%
 }%
 {%
   \mathit{\tuplec{imem}{pc+1}{stk}}%
 }%
 \Rule{%
   \mathit{\tuplec{imem}{pc}{stk} : imem[pc] = \langle nop \rangle}%
  }%
  {%
    \mathit{\tuplec{imem}{pc+1}{stk}}%
  }%
 \Rule{%
   \mathit{\tuplec{imem}{pc}{stk} : imem[pc] =  nil }%
  }%
  {%
    \mathit{\tuplec{imem}{pc+1}{stk}}%
  }%
\end{minipage}}
\hfill
\fbox{\begin{minipage}{.97\textwidth} BSTK
  ($\xrightarrow{C}$) where $k=$ capacity of \mmit{ibuf}, $m=\mmit{|ibuf|}$ \newline%
  \Rule{%
    \mathit{\qtuplec{imem}{pc}{ibuf}{stk} : m < k, \;  
      imem[pc] \neq \La top \Ra }%
  }%
  {%
    \mathit{\qtuplec{imem}{pc+1}{ibuf\,{::}\, imem[pc]}{stk}}%
  }%
  \Rule{%
    \hspace*{-8pt}\mathit{\qtuplec{imem}{pc}{ibuf}{stk} : imem[pc] = \La top \Ra}, \\%
    \mathit{\tuplec{ibuf}{0}{stk}
      \xrex{A}\hspace{-1.5pt}\vspace{-2pt}^m \tuplec{ibuf}{m}{stk'}}%
  }%
  {%
    \mathit{\qtuplec{imem}{pc+1}{[]}{stk'}}%
  }%
  \Rule{%
    \hspace*{-8pt}\mathit{\qtuplec{imem}{pc}{ibuf}{stk} : imem[pc] = nil}, \\%
    \mathit{\tuplec{ibuf}{0}{stk}
      \xrex{A}\hspace{-1.5pt}\vspace{-2pt}^m \tuplec{ibuf}{m}{stk'}}%
  }%
  {%
    \mathit{\qtuplec{imem}{pc+1}{[]}{stk'}}%
  }%
  \Rule{%
    \hspace*{-8pt}\mathit{\qtuplec{imem}{pc}{ibuf}{stk} : m = k}, \\%
    \mathit{\tuplec{ibuf}{0}{stk}
      \xrex{A}\hspace{-1.5pt}\vspace{-2pt}^m \tuplec{ibuf}{m}{stk'}}%
  }%
  {%
    \mathit{\qtuplec{imem}{pc+1}{[imem[pc]]}{stk'}}%
  }%
\end{minipage}
}
\caption{\footnotesize{Syntax and Semantics of Stack and Buffered
    Stack Machine}}
\label{fig:stacksemantics}
\end{figure}

STK fetches an instruction from the instruction memory, executes it,
increases the program counter, and possibly modifies the stack, as
outlined in \autoref{fig:stacksemantics}.  

STK fetches an instruction from the \p{imem}, executes it, increments
the \p{pc} by 1, and possibly modifies the \p{stk}, as outlined in
\autoref{fig:stacksemantics}. Since STK is a deterministic machine, we
formalize its transition relation using a function \p{spec-step},
which uses an auxiliary function \p{stk-step-inst} to capture the
effect of executing an instruction on the stack. We are now ready to
define the transition system \M{A} of STK machine. The set of states
$S_A$ in the transition system \M{A} is is the set of all states
satisfying the predicate \p{sstatep}. Two states $s,u \in S_A$ are
related by transition relation $\xrightarrow{A}$ iff it is possible in
one step to transition from $s$ to $u$, \ie \p{u = (spec-step s)}. The
labeling function, $L_A$ is the identity function.
\begin{alltt}
(defun stk-step-inst (inst stk)
 "returns next state of stk"
  (let ((op (car inst)))
    (cond ((equal op 'push)
           (mpush (cadr inst) stk ))
          ((equal op 'pop)
           (mpop stk))
          ((equal op 'top)
           (mtop stk))
          (t stk))))

(defun spec-step (s)
  "single step of STK machine"
  (let* ((pc (sstate-pc s))
         (imem (sstate-imem s))
         (inst (nth pc imem))
         (stk (sstate-stk s)))
    (if (instp inst)
        (sstate imem (1+ pc) (stk-step-inst inst stk))
      (sstate imem (1+ pc) stk))))
\end{alltt}

The state of BSTK is similar to STK, except that it also includes an
instruction buffer \p{ibuf}. The instruction buffer is encoded as a
list of instructions with an additional restriction on its capacity
\phantom{ } {(\p{ibuf-capacity})}. To encode \p{ibuf} in the
\p{defdata} framework, we have at least two choices. We can use the
\p{oneof} \p{defdata} construct to encode it as an empty list or list
of one, two, or three instructions. Another way is to use the
capability of the \p{defdata} framework to define custom data
types. In the later case, we first define a recognizer function
\p{inst-buffp} and an enumerator function \p{nth-inst-buff}.
\begin{alltt}
(defun inst-buffp (l)
  (and (inst-memp l)
       (<= (len l) (ibuf-capacity))))

(defun nth-inst-buff (n)
  (let ((imem (nth-inst-mem n)))
    (if (<= (len imem) (ibuf-capacity))
        imem
      (let ((i1 (car imem))
            (i2 (cadr imem))
            (i3 (caddr imem)))
        (list i1 i2 i3)))))
\end{alltt}

\noindent We can now register our custom type \p{inst-buff} using the
\p{register-custom-type} macro. Once we have registered it as a
\p{defdata} type we can use it just like other type directly introduced
using \p{defdata} construct.
\begin{alltt}
(register-custom-type inst-buff :enumerator nth-inst-buff 
                                :predicate inst-buffp)
\end{alltt}

\noindent We can now define state of BSTK machine using \p{defdata}
\p{record} construct.

\begin{alltt}
(defdata istate
  (record (imem . inst-mem)
          (pc . nat)
          (stk . stack)
          (ibuf . inst-buff)))
\end{alltt}

BSTK fetches an instruction from the instruction memory, and if the
instruction fetched is not \mmit{top} and the instruction buffer is
not full (function \emph{stutterp} below), it queues the fetched
instruction to the end of the instruction buffer and increments the
program counter. If the instruction buffer is full, then the machine
executes all buffered instructions in the order they were enqueued,
thereby draining the instruction buffer and obtaining a new stack. It
also updates the instruction buffer so that it only contains just the
current fetched instruction. If none of the transition rules match,
then BSTK drains the instruction buffer (if it is not empty) and
updates the stack accordingly. Since BSTK is also a deterministic
machine, we encode its transition relation ($\xrightarrow{C}$) as the
function \p{impl-step}. Having defined the transition relation and the
state of BSTK machine, we can define its transition system \M{C}.

\begin{alltt}
(defun stutterp (inst ibuf)
  "BSTK stutters if ibuf is not full or the current instruction is not 'top"
  (and (< (len ibuf) (ibuf-capacity))
       (not (equal (car inst) 'top))))

(defun impl-step (s)
  "single step of BSTK"
  (let* ((stk (istate-stk s))
         (ibuf (istate-ibuf s))
         (imem (istate-imem s))
         (pc (istate-pc s))
         (inst (nth pc imem)))
    (if (instp inst)
        (let ((nxt-pc (1+ pc))
              (nxt-stk (if (stutterp inst ibuf)
                           stk
                         (impl-observable-stk-step stk ibuf)))
              (nxt-ibuf (if (stutterp inst ibuf)
                            (impl-internal-ibuf-step inst ibuf)
                          (impl-observable-ibuf-step inst))))
          (istate imem nxt-pc nxt-stk nxt-ibuf))
      (let ((nxt-pc (1+ pc))
            (nxt-stk (impl-observable-stk-step stk ibuf))
            (nxt-ibuf nil))
        (istate imem nxt-pc nxt-stk nxt-ibuf)))))
\end{alltt}


Before we describe the correctness of BSTK based on skipping
refinement, we first discuss why an existing notion of refinement such
as stuttering refinement~\cite{manolios2001mechanical} will not
suffice. If BSTK takes a step, which requires it to drain its
instruction buffer (the buffer is full or the current instruction
fetched is \p{top}), then the stack will be updated to reflect the
execution of all instructions in \p{ibuf}, something that is neither a
stuttering step nor a single transition of the STK system. Therefore,
it is not possible to prove that BSTK refines STK, using stuttering
refinement and a refinement map that does not transform the \p{stack}.

We now formulate the correctness of BSTK based on the notion of
skipping refinement. We show \M{C} $\lesssim_r$ \M{A}, using
\autoref{def:wfsk}. We define the refinement map, but first we note
that we do not have to consider all syntactically well-formed STK
states. We only have to consider states whose instruction buffer is
consistent with the contents of the instruction memory, so called
\emph{good states}~\cite{manolios2005computationally}. One way of
defining a good state is as follows: state $s$ is good iff $\mmit{pc
\geq |ibuf|}$ and stepping BSTK from $\mmit{\langle imem, pc - |ibuf|,
[], stk \rangle}$ state for $\mmit{|ibuf|}$ steps yields state $s$,
where $\mathit{|ibuf|}$ is number of instructions in the instruction
buffer of state $s$. We define a predicate \p{good-statep} recognizing
a good state and show that the set of good states is closed under the
transition relation of BSTK.
\begin{alltt}
(defun commited-state (s)
  (let* ((stk (istate-stk s))
         (imem (istate-imem s))
         (ibuf (istate-ibuf s))
         (pc (istate-pc s))
         (cpc (nfix (- pc (len ibuf)))))
    (istate imem cpc stk nil)))

(defun good-statep (s)
  "if state s is reachable from a commited-state in |ibuf| steps"
  (let ((pc (istate-pc s))
        (ibuf (istate-ibuf s)))
    (and (istatep s)
         (>= pc (len ibuf))
         (let* ((cms (commited-state s))
                (s-cms (cond ((endp ibuf)
                              cms)
                             ((endp (cdr ibuf))
                              (impl-step cms))
                             ((endp (cddr ibuf))
                              (impl-step (impl-step cms)))
                             ((endp (cdddr ibuf))
                              (impl-step (impl-step (impl-step cms))))
                             (t cms))))
           (equal s-cms s)))))

(defthm good-state-inductive
  (implies (good-statep s)
           (good-statep (impl-step s))))
\end{alltt}

The refinement map $\p{ref-map}$, a function from a set of good states
to set of abstract states (\p{sstatep}) is defined as follows.

\begin{alltt}

(defun ref-map (s)
  (let* ((stk (istate-stk s))
         (imem (istate-imem s))
         (pc (istate-pc s))
         (ibuf (istate-ibuf s))
         (ibuflen (len ibuf))
         (rpc (nfix (- pc ibuflen))))
    (sstate imem rpc stk)))
\end{alltt}

Given \p{ref-map}, we define $B$ to be the binary relation induced by
it, \ie $sBw$ iff $s$ is a good state and $w = \p{ref-map}(s)$.

Now observe that when the instruction is full or the current
instruction is \mmit{top}, one step of BSTK corresponds to largest
number of STK steps. In both cases, the BSTK machine executes all
instructions in the instruction buffer and if the current instruction
is \mmit{top}, it executes it as well. The condition WFSK2d in
\autoref{def:wfsk} that requires us to reason about reachability,
hence can easily be reduced to bounded reachability. Hence, we set $j
= k+2$, where $k$ is the capacity of the instruction buffer, and
condition WFSK2d is $\La \exists v : w \trans^{< j} v: uBv \Ra$.

Since STK and BSTK are deterministic machines and STK does not
stutter, we only need to define one rank function, a function from set
of good states to non-negative integers.

\begin{alltt}
(defun rank (s)
  "rank of an istate s is capacity of ibuf - |ibuf|"
   (- (ibuf-capacity) (len (istate-ibuf s))))
\end{alltt}

\noindent With above observations we simplify WFSK2
(\autoref{def:wfsk}) to following condition.

\noindent For all \p{s,u} such that \p{s} and \p{u} are good states
and \p{u = (ref-map s)}
\begin{flalign*}
  \label{eq:sksforstk}  
  &\p{(ref-map s)} \sxrex{A}{k+2} \p{(ref-map u)} \ \vee &\\
  &(\p{(ref-map u)} = \p{(ref-map s)} \wedge \p{(rank u)} < \p{(rank s)})& \qquad \qquad \qquad (1)
\end{flalign*}

Notice that since BSTK is deterministic, $u$ is a function of $s$, so
we can remove $u$ from the above formula. Since $k+2$ is a constant,
we can expand out $\sxrex{A}{k+2}$ using only $\xrightarrow{A}$
instead. We formalize Equation 1 in ACL2s by first defining a function
\p{spec-step-skip-rel}, which takes as input STK states $v$ and $w$
and returns true only if $v$ is reachable from $w$ in
\p{(ibuf-capacity)} + 1 steps.


\begin{alltt}
(defthm bstk-skip-refines-stk
  (implies (and (good-statep s)
                (equal w (ref-map s))
                (equal u (impl-step s))
                (not (and (equal w (ref-map u))
                          (< (rank u) (rank s)))))
           (spec-step-skip-rel w (ref-map u))))
\end{alltt}

Once the definitions were in place, proving \p{bstk-skip-refines-stk}
with ACL2s was straightforward. Next, we evaluated how amenable is SKS
for automated reasoning, \ie using \emph{only} symbolic simulation and
no additional lemmas. We model BSTK with instruction buffer capacity
of 2, 3, and 4. while no other restrictions were placed on the
machines.  In particular, the instruction memory (\p{imem}) and the
stack (\p{stk}) component of the state for BSTK and STK machines are
unbounded. The experiments were run on a 2.2 GHz Intel Core i7 with 16
GB of memory. For the BSTK with instruction buffer capacity of 2
instructions, it took $\sim 12$ minutes to complete the proof and for
a BSTK with instruction buffer capacity of 3 instructions, it took
$\sim 2$ hours. For BSTK with instruction buffer capacity of 4
instructions the proof did not finish in over 3 hours.   

\subsection{Memory Controller}
\label{sec:memc}

A memory controller is an interface between a CPU and a memory, and
synchronizes communication between them. Designers implement several
optimizations in a memory controller to maximize available memory
bandwidth utilization and reduce the latency of memory accesses, known
bottlenecks in optimal performance of programs. In this case study, we
consider OptMEMC, a simple model of such an optimized memory
controller. In our simplified model, a CPU is modeled as a list of
memory request (\mmit{reqs}) and memory as a list of natural
numbers (\mmit{mem}).

\begin{figure}[p!]
 \begin{flalign*}
   &\mathit{mem} := \mathit{[] \vbar v\,{::}\,mem}&\text{(Memory)}&\\
   &\mathit{req} := \mathit{\langle write\ addr\ v \rangle\vbar
     \langle read\ addr \rangle } \vbar \langle refresh \rangle &\text{(Request)}&\\
   &\mathit{pt}:= 0 \vbar 1 \vbar \cdots \vbar n \vbar \cdots &\text{(Request Location)}&\\
   &\mathit{reqs} := \mathit{[] \vbar req\,{::}\,reqs}&\text{(Requests)}&\\
   & \mathit{rbuf} := \mathit{[req_1, \ldots , req_k]}&\text{(Request Buffer)}&\\
   &\mathit{sstate} := \mathit{\tuplec{reqs}{pt}{mem}}&\text{(MEMC State)}&\\
   &\mathit{istate} := \mathit{\langle reqs, pt, rbuf, mem\rangle}
   &\text{(OptMEMC State)}&\qquad
 \end{flalign*}
\hfill
\fbox{\begin{minipage}{\textwidth} MEMC ($\xrightarrow{A}$)%
    \newline%
    \Rule{%
      \mathit{\tuplec{reqs}{pt}{mem}, \; reqs[pt] = \langle write \
        addr\ v \rangle}%
    }%
    {%
      \mathit{\tuplec{reqs}{pt+1}{mem[addr] \leftarrow v}}%
    }%
    \Rule{%
      \mathit{\tuplec{reqs}{pt}{mem}, \; reqs[pt] = \langle read\
        addr\rangle}%
    }%
    {%
      \mathit{\tuplec{reqs}{pt+1}{mem}}%
    }%
    \Rule{%
      \mathit{\tuplec{reqs}{pt}{mem}, \; reqs[pt] = \langle refresh \rangle}%
    }%
    {%
      \mathit{\tuplec{reqs}{pt+1}{mem}}%
    }%
  \end{minipage}}
\fbox{\begin{minipage}[t]{.995\textwidth} OptMEMC
    ($\xrightarrow{C}$)\newline%
    Let $\mathit{|rbuf| = j}$\newline%
    \Rule{%
      \mathit{\qtuplec{reqs}{pt}{rbuf}{mem},\quad j < k},
      \mathit{req \neq top }%
    }%
    {%
      \mathit{\qtuplec{reqs}{pt}{rbuf\,{::}\, reqs[pt]}{mem}}%
    }%
    \Rule{%
      \mathit{\qtuplec{reqs}{pt}{rbuf}{mem}, \quad reqs[pt] = \langle
        read\ addr \rangle}, \\%
      \mathit{\tuplec{rbuf}{0}{mem}
        \xrex{A}\hspace{-1.5pt}\vspace{-2pt}^j \tuplec{rbuf}{j}{mem'}}%
    }%
    {%
      \mathit{\qtuplec{reqs}{pt}{[]}{mem'}}%
    }%
    \Rule{%
      \mathit{\qtuplec{reqs}{pt}{rbuf}{mem},\quad j = k}, \\
      \mathit{\tuplec{rbuf}{0}{mem}
        \xrex{A}\hspace{-1.5pt}\vspace{-2pt}^j \tuplec{rbuf}{k}{mem'}}%
    }%
    {%
      \mathit{\qtuplec{reqs}{pt}{rbuf\,{::}\, reqs[pt]}{mem'}}%
    }%
  \end{minipage}
}
\caption{\footnotesize{Syntax and Semantics of MEMC and OptMEMC}}
\label{fig:memsemantics}
\end{figure}

OptMEMC fetches a memory request from location \mmit{pt} in a queue of
CPU requests, \mmit{reqs}. It enqueues the fetched request in the
request buffer, $\mathit{rbuf}$ and increments \mmit{pt} to point to
the next CPU request in \mmit{reqs}. The capacity of \mmit{rbuf} is
$k$, a fixed positive integer. If the fetched request is a
\mmit{read} or the request buffer is full, then before enqueuing the
request into \mmit{rbuf}, OptMEMC first analyzes the request buffer
for consecutive write requests to the same address in the memory
(\mmit{mem}). If such a pair of writes exists in the buffer, it marks
the older write requests in the request buffer as redundant. Then it
executes all the requests in the request buffer except the one that
are marked redundant. Requests in the buffer are executed in the order
they were enqueued. In addition to read and write commands, the memory
controller periodically issues a \mmit{refresh} command to preserve
data in memory. A \mmit{refresh} command reads all memory locations
and immediately writes them back without modification. Refresh
commands are required to periodically reinforce the charge in the
capacitive storage cells in a DRAM. In effect, a refresh command
leaves the data memory unchanged. We define the function \p{mrefresh}
and prove that the memory is same before and after execution of the
\p{refresh} command. This is the only property of \p{mrefresh} that we
would require.

\begin{alltt}
(defthm mrefresh-mem-unchanged
  (equal (mrefresh mem)
         mem))
\end{alltt}

To reason about the correctness of OptMEMC using skipping refinement,
we define a high-level abstract system, MEMC, that acts as the
specification for OPTMEMC. It fetches a memory request from the CPU
and immediately executes the request. The syntax and the semantics of
MEMC and OPTMEMC are given in~\autoref{fig:memsemantics}, using the
same conventions as described previously in the stack machine section.

We now formulate the correctness of OptMEMC based on the notion of
skipping refinement. Let \trs{A} and \trs{C} be transition systems for
MEMC and OptMEMC respectively. Like in the previous case study, we
encode the state of the machines using \p{defdata} and formalize the
transition relation of OptMEMC and MEMC using a step function that
describes the effect of each instruction on the state of the
machine. The labeling function $L_A$ and $L_C$ are the identity
functions. Given a refinement map $\p{ref-map} : S_C \rightarrow S_A$,
we use~\autoref{def:wfsk} to show that $\M{C} \lesssim_r \M{A}$. As
was the case with the previous case study, OptMEMC and MEMC are
deterministic machines and MEMC does not stutter.  WFSK2
(\autoref{def:wfsk}) can again be simplified to Formula 1.

Once the definitions of the transition systems for the two machines
were in place, it was straightforward to prove skipping refinement
with ACL2s. Like in the previous case study, we also prove the theorem
using \emph{only} symbolic execution and no additional lemmas, for
configurations of OPTMEMC with buffer capacity of 2 and 3. For OptMEMC
with buffer capacity of 2, the final theorem was proved in $\sim 2$
minutes and with OptMEMC buffer capacity of 3, it took $\sim 1$ hour
to prove the final theorem. The proof with buffer capacity of 4
instructions did not finish in over 3 hours.

\subsection{Superword Level Parallelism with SIMD instructions}
\label{sec:vectorize}
An effective way to improve the performance of multimedia programs
running on modern commodity architectures is to exploit
Single-Instruction Multiple-Data (SIMD) instructions (\eg the SSE/AVX
instructions in x86 microprocessors). Compilers analyze programs for
\emph{superword level parallelism} and when possible replace multiple
scalar instructions with a compact SIMD instruction that concurrently
operates on multiple data \cite{larsen2000exploiting}. In this case
study, we illustrate the applicability of skipping refinement to
verify the correctness of such a compiler transformation.

For the purpose of this case study, we make some simplifying
assumptions: the state of the source and target programs (modeled as
transition systems) is a three-tuple consisting of a sequence of
instructions, a program counter and a store. We also assume that a
SIMD instruction simultaneously operates on \emph{two} sets of data
operands and that the transformation analyzes the program at a basic
block level. Therefore, we do not model any control flow
instruction. \autoref{fig:superword} shows how two add and two
multiply scalar instructions are transformed into corresponding SIMD
instructions. Notice that the transformation does \emph{not} reorder
instructions in the source program.

\begin{figure}[h]
\hspace{4cm}
   \begin{minipage}[t][1.3cm][t]{3cm}
     \begin{tabular}{ccccc}
       a & = & b & + & c\\
       d & = & e & + & f\\
     \end{tabular}
   \end{minipage}
   $\rightarrow$\quad
   \begin{minipage}[t]{\textwidth}
     \fbox{\begin{minipage}[c][.7cm][t]{3pt}
         a
         d
       \end{minipage}
     }
     =
     \fbox{\begin{minipage}[c][.7cm][t]{3pt}
         b
         e
       \end{minipage}
     }
     ${+_{\tiny{SIMD}}}$
     \fbox{\begin{minipage}[c][.7cm][t]{3pt}
         c
         f
       \end{minipage}
     }
   \end{minipage}

\hspace{4cm}
   \begin{minipage}[b]{3cm}
     \begin{tabular}{ccccc}
       u & = & v & $\times$ & w \\
       x & = & y & $\times$ & z\\
     \end{tabular} 
   \end{minipage}     
   $\rightarrow$\quad
   \fbox{\begin{minipage}[c][.7cm][t]{3pt}
       u
       x
     \end{minipage}
   }
   =
   \fbox{\begin{minipage}[c][.7cm][t]{3pt}
       v
       y
     \end{minipage}
   }
   ${\times_{\tiny{SIMD}}}$
   \fbox{\begin{minipage}[c][.7cm][t]{3pt}
       w
       z
     \end{minipage}
   }
   \caption{Superword Parallelism}
   \label{fig:superword}
 \end{figure}

 The syntax and operational semantics of the scalar and vector
 machines are given in~\autoref{fig:scalarmergesemantics}, using the
 same conventions as described previously in the stack machine
 section. We denote that $x, \ldots, y$ are variables with values
 $v_x, \ldots, v_y$ in \mmit{store} by $\{ \langle x, v_x \rangle,
 \ldots, \langle y, v_y \rangle \} \subseteq \mathit{store}$. We use
 $\mathit{\sem{(sop \ v_x\ v_y)}}$ to denote the result of a scalar
 operation $\mathit{sop}$ and $\mathit{\sem{(vop \ \langle v_a \ v_b
 \rangle \langle v_d \ v_e \rangle)}}$ to denote the result of a
 vector operation $\mathit{vop}$. Finally, we use $\mmit{store}|_{x :=
 v_x, \ldots, y := v_y}$ to denote that variables $x, \ldots, y$ are
 updated (or added) to \mmit{store} with values $v_x, \ldots,
 v_y$. Notice that the language of a source program consists of scalar
 instructions while the language of the target program consists of
 both scalar and vector instructions. As in the previous two case
 studies, we model the transition relation of a program (both source
 and target program) by modeling the effect of an instruction on the
 state of machines.


\begin{figure}[h]
   \begin{flalign*}
     &\mathit{loc} := \mathit{\{x,y,z,a,b,c, \ldots\}}&\text{(Variables)}&\\
     &\mathit{sop} := \mathit{add} \vbar \mathit{sub}
     \vbar \mathit{mul} \vbar \mathit{and} \vbar
     \mathit{or} \vbar \mathit{nop} &\text{(Scalar Ops)}&\\
     &\mathit{vop} := \mathit{vadd} \vbar
     \mathit{vsub} \vbar \mathit{vmul} \vbar \mathit{vand}
     \vbar \mathit{vor} \vbar \mathit{vnop} &\text{(Vector Ops)}&\\
     &\mathit{sinst} := \mathit{sop \tuple{z}{x}{y}}
     &\text{(Scalar Inst)}&\\
     &\mathit{vinst} := \mathit{vop\tuple{c}{a}{b} \tuple{f}{d}{e}}
     &\text{(Vector Inst)}&\\
     &\mathit{sprg} := \mathit{[] \vbar sinst\,{::}\,sprg}&
     \text{(Scalar Program)}&\\
     &\mathit{vprg} := \mathit{[] \vbar (sinst \vbar vinst)\,{::}\,vprg}&
     \text{(Vector Program)}&\\
     &\mathit{store} := \mathit{[] \vbar \langle x, v_x \rangle
       \,{::}\,store} &\text{(Registers)}
  \end{flalign*}
  \hfill
\fbox{\begin{minipage}{.978\textwidth}
  \textrm{Scalar Machine ($\xrightarrow{A}$)}
  \vspace{-10pt}
  \Rule{
    \mathit{\langle sprg, pc, store\rangle, \; 
      \{ \langle x, v_x\rangle,  \langle y, v_y\rangle \} \subseteq store,}\\
  \mathit{{sprg[pc] = sop \tuple{z}{x}{y}},\;
    {v_z = \sem{(sop \ v_x\ v_y)}}}
    }
    {
      \mathit{\tuplec{sprg}{pc+1}{store|_{z :=  v_z}}}
    }
  \end{minipage}
}
\hfill
\fbox{
\begin{minipage}{.97\textwidth}
  \textrm{Vector Machine ($\xrightarrow{C}$)}
  \vspace{-10pt}
  \Rule{
    \mathit{{\langle vprg, pc, store\rangle}, \;
      \{\langle x, v_x\rangle, \langle y, v_y\rangle\} \subseteq
    store,}\\
  \mathit{{sprg[pc] = sop \tuple{z}{x}{y}},\;
    {v_z = \sem{(sop \ v_x\ v_y)}}}
  }
  {
    \mathit{\tuplec{vprg}{pc+1}{store|_{z :=  v_z}}}
  }
  \Rule{%
    \mathit{{\tuplec{vprg}{pc}{store}},\; {vprg[pc] = vop 
        \tuple{c}{a}{b} \tuple{f}{d}{e}}} \comment{\quad \textbf{c
        $\neq$ d,e}},\\
    \mathit{\{\langle a, v_a\rangle , \langle b , v_b\rangle , \langle d
        , v_d\rangle , \langle e ,  v_e\rangle\} \subseteq store},\\%
    \mathit{\langle v_c, v_f\rangle = \sem{(vop \
        \langle v_a \ v_b \rangle \langle v_d \ v_e \rangle)}}\\%
  }%
  {%
    \mathit{\tuplec{vprg}{pc+1}{store|_{c :=  v_c, f:=v_f }}}
  }
\end{minipage}
}
\caption{Syntax and Semantics of Scalar and Vector Program}
    \label{fig:scalarmergesemantics}
\end{figure}
\vspace{-.1cm}

We use the translation validation approach to verify the correctness
of the vectorizing compiler transformation~\cite{barrett2005tvoc}, \ie
we prove the equivalence between a source program and the generated
vector program. As in the previous two case studies, the notion of
stuttering simulation is too strong to relate a scalar program and the
vector program produced by the vectorizing compiler, no matter what
refinement map we use. To see this, note that the vector machine might
run exactly twice as fast as the scalar machine and during each step
the scalar machine might be modifying the memory. Since both machines
do not stutter, in order to use stuttering refinement, the length of
the vector machine run has to be equal to the run of the scalar
machine.

Let \trs{A} and \trs{C} be transition systems of the scalar and vector
machines, respectively corresponding to the source and target
programs. The vector program is correct iff \M{C} refines \M{A}.  We
show $\M{C} \lesssim_r \M{A}$, using~\autoref{def:wfsk}.  Determining
$j$, an upper-bound on skipping that reduces condition WFSK2d
in~\autoref{def:wfsk} to bounded reachability is simple because the
vector machine can perform at most 2 steps of the scalar machine at
once; therefore $j=3$ suffices.

We next define the refinement map. Recall that refinement maps are
used to define what is observable at concrete states from viewpoint of
the abstract system. Let \mmit{sprg} be the source program and
\mmit{vprg} be the compiled vector program. We first define a function
\p{pcT} that takes as input the vector machine's program counter
\p{pc} and a vector program \p{vprg} and returns the corresponding
value of the scalar machine's program counter.
\begin{alltt}
(defun num-scaler-inst (inst)
  (cond ((vecinstp inst)
         2)
        ((instp inst)
         1)
        (t 0)))

(defun pcT (pc vprg)
"maps values of the vector machine's pc to the corresponding values of
the scalar machine's pc"
  (let ((inst (nth pc vprg)))
   (cond ((or (not (integerp pc))
              (< pc 0))
          0)
         ((zp pc)
          (num-scaler-inst inst))
         (t (+ (num-scaler-inst inst) (pcT (1- pc) vprg))))))
\end{alltt}

We next define a function \p{scalarize-vprg} that takes as input a
vector program \p{vprg}. It walks through the list of instructions in
\p{vprg} and translates each instruction in one the following ways: if
it is a vector instruction it \emph{scalarizes} it into a list of
corresponding scalar instructions, else if it is a scalar instruction
it returns the list containing the instruction itself (function
\p{scalarize} below). The result of \p{scalarize-vprg} is a scalar
program. Notice that this function is significantly simpler than the
compiler transformation procedure. This is because the complexity of a
compiler transformation typically lies in its analysis phase, which
determines if the transformation is even feasible, and not in the
transformation phase itself.

\begin{alltt}
(defun scalarize (inst)
"scalerize a vector instruction"
  (cond ((vecinstp inst)
         (let ((op (vecinst-op inst))
               (ra1 (car (vecinst-ra inst)))
               (ra2 (cdr (vecinst-ra inst)))
               (rb1 (car (vecinst-rb inst)))
               (rb2 (cdr (vecinst-rb inst)))
               (rc1 (car (vecinst-rc inst)))
               (rc2 (cdr (vecinst-rc inst))))
          (case op
            (vadd (list (inst 'add rc1 ra1 rb1)
                        (inst 'add rc2 ra2 rb2)))
            (vsub (list (inst 'sub rc1 ra1 rb1)
                        (inst 'sub rc2 ra2 rb2)))
            (vmul (list (inst 'mul rc1 ra1 rb1)
                        (inst 'mul rc2 ra2 rb2))))))
        ((instp inst) (list inst))
        (t nil)))

(defun scalarize-vprg-aux (pc vprg)
"scalerize the vector program from [0,pc]"
  (if (or (not (integerp pc))
          (< pc 0))
      nil
    (let ((inst (nth pc vprg)))
     (cond
      ((zp pc) ;=0
       (scalarize inst))
      (t
       (append (scalarize-vprg (1- pc) vprg) (scalarize inst)))))))

(defun scalarize-vprg (vprg)
 (scalarize-vprg-aux (len vprg) vprg))
\end{alltt}

\noindent The refinement map \p{ref-map}: $S_C \ra S_A$, now can be
defined as follows.

\begin{alltt}
(defun ref-map (s)
  (let* ((store (vstate-store s))
         (vprg (vstate-vprg s))
         (isapc (pcT (1- (vstate-pc s)) vprg)))
    (sstate isapc store (scalarize-vprg (len vprg) vprg))))
\end{alltt}

\noindent Given \p{ref-map}, we define $B$ to be the binary relation
induced by the refinement map, \ie $sBw$ iff $s \in S_C$ and
$w = \p{(ref-map s)}$. Notice that since the machines do not stutter,
WFSK2 (\autoref{def:wfsk}) can be simplified as follows.  For all
$s, u \in S_C$ such that $s \xrightarrow{C} u$:
\begin{flalign*}
  \label{eq:vec-skipping-thm}  
  &\p{(ref-map s)} \sxrex{A}{3} \p{(ref-map u)} & (2)
\end{flalign*}

Since the vector machine is deterministic, $u$ is a function of $s$,
so we can remove $u$ from the above formula, if we wish.  Also, we can
expand out $\sxrex{A}{3}$ to obtain a formula using only
$\xrightarrow{A}$ instead. We prove the appropriate lemmas to prove
the final theorem: vector machine refines scalar machine.
\begin{alltt}
(defthm vprg-skip-refines-sprg
  (implies (and (vstatep s)
                (equal w (ref-map s)))
           (spec-step-skip-rel w (ref-map (vec-step s)))))
\end{alltt}

where \p{vstatep} is the recognizer for a state of vector machine;
\p{vec-step} is a transition function for vector machine; and
\p{spec-step-skip-rel} is a function that takes as input two states of
scalar machine and returns true if the second is reachable from the
first in less than three steps.

Note that \mmit{pcT(pc, vprg)} can also be determined using a history
variable and would be a preferable strategy from verification
efficiency perspective.

\vspace{-10pt}
\section{Conclusion and Future Work}
\label{sec:conclusion}

In this paper, we used skipping refinement to prove the correctness of
three optimized reactive systems in ACL2s. The concrete optimized
systems can run ``faster'' than the corresponding abstract high-level
specifications. Skipping refinement is an appropriate notion of
correctness for reasoning about such optimized systems. Furthermore,
well-founded skipping simulation gives ``local'' proof method that is
amenable for automated reasoning.  Stuttering simulation and
bisimulation have been used widely to prove correctness of several
interesting
systems~\cite{manolios2000correctness,ray2004deductive,ray2013specification}.
However, we have shown that these notions are too strong to analyze
the class of optimized reactive systems studied in this paper.
Skipping simulation is a weaker and more generally applicable notion
than stuttering simulation.  In particular, skipping simulation can be
used to reason about superscalar processors, pipelined processors with
multiple instructions completion, without modifying the specification
(ISA), an open problem in~\cite{ray2004deductive}.  We refer the
reader to our companion paper~\cite{manolios2015sks} for a more
detailed discussion on related work.

For future work, we would like to develop a methodology to increase
proof automation for proving correctness of systems based on skipping
refinement. In~\cite{manolios2015sks}, we showed how model-checkers
can be used to analyze correctness for finite-state systems.
Similarly, we would like to use the GL
framework~\cite{swords2010verified}, a verified framework for symbolic
execution in ACL2, to further increase the efficiency and
automation. 

\vspace{-10pt} {\section*{\smaller{Acknowledgments}} { We would like
    to thank Harsh Raju Chamarthi for help on the proof of vectorizing
    compiler transformation.}}

\bibliographystyle{eptcs}
\bibliography{paper}

\end{document}